\documentclass[twocolumn,showpacs,superscriptaddress,preprintnumbers,amsmath,amssymb,epsfig,floatfix,aps,bibnotes]{revtex4-1}\usepackage{amssymb}

\usepackage{lipsum}
\usepackage{amsmath}
\usepackage{upgreek}
\usepackage{epsfig}
\usepackage{graphicx}
\usepackage{dcolumn}
\usepackage{color}
\usepackage{natbib}  
\usepackage{hyperref}
\usepackage{breakurl}
\usepackage{mathrsfs}
\hypersetup{colorlinks=true, citecolor=blue, urlcolor=blue, linkcolor=blue}
\usepackage{bm}
\usepackage{adjustbox}
\usepackage{tabularx}
\newcolumntype{L}[1]{>{\raggedright\arraybackslash}p{#1}}
\newcolumntype{C}[1]{>{\centering\arraybackslash}p{#1}}
\newcolumntype{R}[1]{>{\raggedleft\arraybackslash}p{#1}}
\usepackage[utf8]{inputenc}
\inputencoding{latin1}
\inputencoding{utf8}
\usepackage[T1]{fontenc}

\usepackage[capitalize]{cleveref}

\begin{document}
\title{Mechanistic insights on the phosphorene degradation}
\author{Rohit Babar}
\affiliation{Department of Physics, Indian Institute of Science Education and Research, Pune 411008, India}	
\author{Mukul Kabir}
\email{Corresponding author: mukul.kabir@iiserpune.ac.in} 
\affiliation{Department of Physics, Indian Institute of Science Education and Research, Pune 411008, India}
\affiliation{Centre for Energy Science, Indian Institute of Science Education and Research, Pune 411008, India}
\date{\today}

\begin{abstract} 
The structural and chemical degradations of phosphorene severely limit its practical applications despite the enormous promise. In this regard, we investigate a plethora of microscopic kinetic mechanisms and develop a degradation phase diagram within the first-principles calculations.  At 400 K, the degradation and the competing self-annealing proceeds through the merger and annihilation of vacancies, respectively, which are triggered via itinerant vacancy and adatom.  A further increase in temperature beyond 650 K, the structural degradation results through the emission of the undercoordinated atoms from the defect and the concurrent pair-wise sublimation. The role of  inter-layer vacancy diffusion is discarded in the context of structural degradation. The chemical degradation is routed through the dissociation of oxygen molecule that is either activated at the room-temperature on the pristine surface or spontaneous at the single-vacancy site.  The present results are in agreement with the few available experimental  conjectures and will motivate further efforts. 
\end{abstract}
\maketitle

\section {Introduction}
Few-layered black phosphorus  (BP) has garnered significant attention due to a layer-tuneable band gap between 0.3--2.0 eV and anisotropic mechanical, electronic and optical responses.~\citep{10.1021/nn501226z,10.1038/nnano.2014.35,Xia2014,10.1021/nn503893j,Luo2015,10.1021/nl502865s,PhysRevLett.112.176801,PhysRevB.89.235319} The staggered hexagonal layers of BP are stacked together by van der Waals interaction, similar to graphite.  The isolated layers of BP, phosphorene, offers significant advantages in realizing novel semiconducting and optoelectronic devices due to its high current on-off ratio, extraordinary carrier mobility, ambipolar transport, and anisotropic electronic dispersion.~\citep{10.1021/nn501226z,10.1038/nnano.2014.35,10.1021/nn505868h,PhysRevB.89.235319,Wang2015,10.1021/nl502892t}  Moreover, the high mechanical flexibility and the retention of extraordinary electronic and optical properties under moderate strain allow to design functional heterostructures with other two-dimensional materials.~\citep{PhysRevLett.112.176801,PhysRevB.90.085402,10.1021/nn505809d,10.1002/anie.201712608} Further, many exotic quantum many-body states have been theoretically predicted and experimentally demonstrated in few-layer phosphorene.~\citep{nl5043769, science.aaa6486,PhysRevB.97.045132} 

However, the major hurdle is the inferior stability of phosphorene at ambient conditions compared to graphene and transition-metal dichalcogenides. The degradation in two-dimension results from the dynamics of lattice defects and chemical interaction with the environment.~\citep{Queisser945,DIEBOLD200353,PhysRevB.76.165202,Song2017} Both surface and subsurface defects have been experimentally observed in the few-layer phosphorene due to their low formation energies, and are responsible for intrinsic p-type conductivity.~\citep{10.1021/nn501226z,10.1021/acs.nanolett.7b00766,10.1063/1.5016988,PhysRevB.91.045433} Similar to graphene, the vacancy defects in phosphorene relaxes in multiple configurations,~\citep{10.1021/acs.jpcc.6b05069,10.1021/jacs.6b04926} which act as anisotropic scatterers and have been further manipulated to form a many-body state with excited carriers.~\citep{10.1021/acs.nanolett.7b00766,10.1021/acs.nanolett.7b03356} In addition to the disruption in the  electronic properties, the vacancies in phosphorene induce mechanical degradation as well as produce structural changes under increasing strain, temperature, or vacancy concentration.~\citep{10.1021/acs.jpclett.5b00043,0957-4484-27-31-315704,10.1063/1.4966167} Vacancy containing strained phosphorene reduces the fracture strength and structural integrity along the zigzag axis due to bond distortion and breakage. Further, regions with high defect density are predicted to initiate crack formation followed by the structural failure under the transverse force generated by an atomic force microscopy nanoindentation.~\citep{10.1021/acs.jpcc.6b13071}   

Thermally activated diffusion of point-defects holds the key to structural changes. Vacancy transformation, diffusion, and aggregation leading to the formation of complex defects and grain boundaries have been extensively studied in graphene,~\citep{C4CS00499J, PhysRevB.98.075439} where the weak bonding at the edges and vacancies initiate structural degradations. The few-layer phosphorene is observed to undergo anisotropic degradation, amorphization, and sublimation when heated above 650 K.~\citep{10.1021/acs.jpclett.5b00043,0957-4484-29-6-065702,10.1021/acs.jpclett.6b00584} However, the current description of structural degradation stands at the diffusion of isolated vacancy within a single layer,~\citep{0957-4484-26-6-065705,10.1021/jacs.6b04926,C6NR05414E} which severely limits our microscopic understanding.  In this regard, here we investigate a cohort of mechanisms involving complex defect-defect interaction, vacancy and adatom migration, surficial desorption, which lead to larger defect formation. The competing self-healing mechanisms of defects are also investigated. Such a comprehensive understanding of degradation is necessary to identify the optimal experimental conditions to prepare and restore the defect-free phosphorene to improve its ambient stability. 

\begin{figure*}[!t]
\begin{center}
\includegraphics[width=0.90\textwidth]{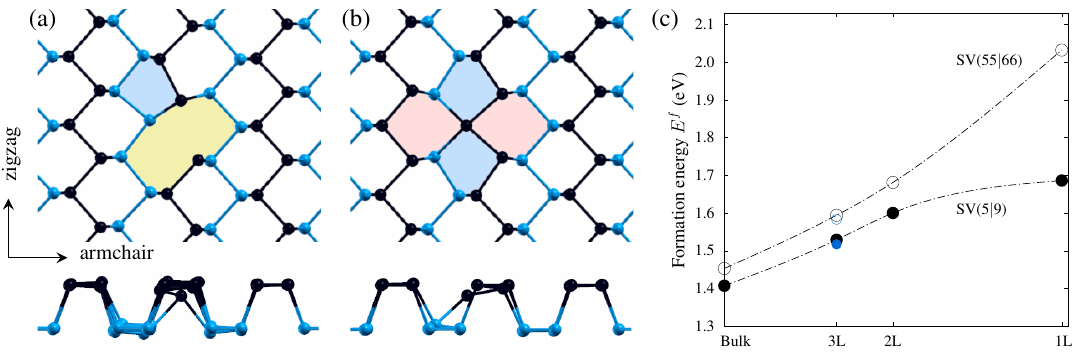}
\caption{The top and side views of the single-vacancy in phosphorene in its two possible configurations (a) SV(5$|$9) and (b) SV(55$|$66). Atoms in the two half-layers are coloured in blue and black. The polygons in the vacancy are shaded with different colours. (c) Formation energies $E^f$ of the SV(5$|$9) and SV(55$|$66) defects at the apical layer with varied layer thickness. The $E^f$ for the subsurface vacancy (blue circles) and bulk are also shown. The formation energy decreases with increasing layer thickness. While in single-layer, the SV(5$|$9) is thermodynamically favourable over the SV(55$|$66)configuration by 350 meV, and the difference in energy between the defects also decreases with the thickness to 45 meV for the bulk black phosphorus. Note that the vacancy formation is much easier in the few-layer phosphorene owing to the much lower $E^f$ compared to graphene and other mono-elemental bulk semiconductors.~\citep{PhysRevB.98.075439,PhysRevB.88.195204,PhysRevB.87.035203} 
}
\label{fig:Fig1config}
\end{center}
\end{figure*}

Rapid oxidation of phosphorene in the ambient environment presents another significant challenge towards its stability. Although the exact mechanism continues to be investigated, it is understood that the strong chemical bonding between lone pair electrons of phosphorus and $\pi^*$ electrons of oxygen initiate the chemical degradation.~\citep{2053-1583-1-2-025001,Favron2015,10.1021/acs.chemmater.6b03592,PhysRevLett.114.046801,10.1021/jacs.8b02156,10.1021/jacs.7b04971,10.1021/acsami.6b16111,PhysRevB.92.125412,10.1002/anie.201705012,10.1002/anie.201605168} Further, degradation is accelerated as the hydrophilicity of phosphorene increases with oxidation.~\citep{2053-1583-1-2-025001,PhysRevLett.114.046801,10.1002/anie.201605168}  Moreover, the edges and steps undergo rapid degradation in few-layered samples, where the role of vacancy defects is understandably negligible.~\citep{10.1021/jacs.8b02156,10.1021/acsami.6b16111} It is also observed that the oxidation rate increases with the reduction in the number of layers,~\citep{Favron2015,10.1021/jacs.7b04971} which indicates a stronger interaction of oxygen at the surface. 

However, the nature of oxygen-defect interaction in phosphorene is a subject of debate at present without any experimental understanding. While the oxidation is proposed to be an activated process, it is more favourable at a lattice vacancy than the pristine surface.~\citep{2053-1583-4-1-015010} On the contrary, a weak interaction was suggested as the vacancy diffusion is found to be unaffected by oxygen.~\citep{10.1021/jacs.6b04926} In the present study, we thoroughly investigate the interaction of O$_2$ on the pristine and defected phosphorene. Further, the microscopic dissociation mechanism is elucidated. We find the O$_2$-dissociation to be a barrier-less process at the single vacancy defect, whereas oxidation on both pristine surface and at the divacancy defect remains an activated process. Once the dissociation is complete, the strong P$-$O binding essentially makes O-removal from the lattice impossible.

\section{Computational details}
The spin-polarized density functional theory calculations are performed within the projector augmented wave formalism,~\citep{PhysRevB.50.17953} as implemented in the Vienna ab initio simulation package (VASP).~\citep{PhysRevB.47.558,PhysRevB.54.11169} The wave functions are described in the plane wave basis with 500 eV cutoff for the kinetic energy, and the exchange-correlation energy is expressed with the Perdew-Burke-Ernzerhof (PBE) functional.~\citep{PhysRevLett.77.3865} Structural relaxations are performed until all the force components fall below 0.01 eV/\AA~ threshold. Calculations on the single-layer phosphorene (SLP) are performed using two supercells with different sizes, 6 (zigzag) $\times$ 4 (armchair) supercell containing 96 atoms and 9 (zigzag) $\times$ 5 (armchair) supercell containing 180 atoms. The reciprocal space integration for the two supercells was evaluated with 4$\times$4$\times$1 and 3$\times$4$\times$1 k-point grid respectively.  A vacuum of 15 \AA~ was always maintained perpendicular to the surface to minimize the spurious interaction between the periodic images.  
Calculations for the few-layer phosphorene are performed using the 6 $\times$ 4 $\times$ $N$ supercells ($N$ $\leqslant$ 3). The nonlocal van der Waals functional (optB88-vdW) of Langreth and Lundqvist is used throughout the calculations to describe the oxygen-phosphorene and interlayer interactions.~\citep{PhysRevLett.92.246401} The activation barriers are calculated within the climbing image nudged elastic band method formalism.~\citep{1.1329672} The true nature of the transition states is confirmed by obtaining the single imaginary vibrational mode. 

\begin{table*}[!]
\caption{Formation energies $E^f$ of the various defect complexes in SLP. The $E^f$ is calculated as, $E^f = E_{\rm tot} - N_{\rm P} \times E_{\rm P}$, where $ E_{\rm tot} $ is the total binding energy of the defect-containing supercell, and $E_{\rm P}$ is the binding energy per P-atom in the pristine SLP. $N_{\rm P}$ is the number of P-atoms in the defected lattice.  The activation energies $E_a$, for the cohort of microscopic processes leading to the self-healing, degradation, and sublimation are calculated. Note that both vacancy and P-adatom P$_{\rm A}$ are highly itinerant due to lower $E_a$ and the corresponding diffusion are highly anisotropic along the zigzag and armchair directions.   
}
\begin{center}
\begin{tabular}{l*{5}{c}c}
 \hline
 Defect complex & & Formation energy  & & Microscopic process & & Activation barrier  \\ 
 & & $E^f$ (eV) & &  & &  $E_a$ (eV) \\
 
\hline
\hline 
 SV(5$|$9)        &  &     1.68    & & 	SV(5$|$9) $\rightarrow$ SV(5$|$9) & & \multicolumn{1}{c}{0.12 (zigzag), 0.38 (armchair)} \\
 SV(55$|$66)      &  &     2.03    & & 	$-$   & & $-$\\
 DV(5$|$8$|$5)    &  &     1.56    & & 	DV(5$|$8$|$5) $\rightarrow$ SV(5$|$9) + SV(9$|$5) & &  \multicolumn{1}{c}{1.64}  \\
 SV(5$|$9) + SV(9$|$5)        &  &  2.52       & &  SV(5$|$9) + SV(9$|$5)	 $\rightarrow$ DV(5$|$8$|$5) & & \multicolumn{1}{c}{0.68} \\ 
 Single P$_{\rm A}$   &  &     1.43    & & 	P$_{\rm A}$ $\rightarrow$ P$_{\rm A}$ & & \multicolumn{1}{c}{0.19 (zigzag), 1.05 (armchair)}  \\ 
 			      &  &             & & 	P$_{\rm A}$ desorption & & \multicolumn{1}{c}{2.14}  \\ 
 P$_{\rm A}$-pair    &  &     2.01    & & 	P$_{\rm A}$-pair desorption & & \multicolumn{1}{c}{barrierless}  \\ 
 SV(5$|$9) + P$_{\rm I}$    &  &     1.56    & & 	SV(5$|$9) + P$_{\rm I}$ $\rightarrow$ pristine SLP & & \multicolumn{1}{c}{0.69}  \\ 
  			      &  &             & & 	pristine SLP $\rightarrow$ SV(5$|$9) + P$_{\rm I}$ & & \multicolumn{1}{c}{2.25}  \\ 
 SV(55$|$66) + P$_{\rm I}$   &  &     1.93    & & SV(55$|$66) + P$_{\rm I}$ $\rightarrow$ SV(5$|$9) + P$_{\rm I}$ & & \multicolumn{1}{c}{0.10}  \\ 
 DV(5$|$8$|$5) + P$_{\rm A}$    &  &     2.68    & & DV(5$|$8$|$5) + P$_{\rm A}$ $\rightarrow$ SV(5$|$9) & & \multicolumn{1}{c}{0.46}  \\ 
   			      &  &             & & SV(5$|$9)	$\rightarrow$ DV(5$|$8$|$5) + P$_{\rm A}$ & & \multicolumn{1}{c}{1.46}  \\
 \hline
\label{tablepvS1} 
\end{tabular}
\end{center}
\end{table*}

\section{Results and Discussion}
The two half-layers in the hexagonal staggered lattice of SLP are separated by 2.13 \AA~ (Figure~\ref{fig:Fig1config}). The in-plane atoms are arranged in a zigzag fashion with 2.23 \AA~ bonds (zigzag axis), whereas the atoms across the two half-layers have slightly larger bond lengths of 2.27 \AA~(armchair axis). Thus, every P-atom is bonded with two in-plane and one out-of-plane atoms. Given that each P-atom has five valence electrons, three electrons participate in bonding, whereas rest of the  electrons produce a lone-pair charge cloud. The Coulomb repulsion among the  lone-pairs leads to a staggered, anisotropic arrangement of P-atoms with comparatively soft bonding (Figure~\ref{fig:Fig1config}). Thus, Young's modulus for phosphorene is substantially lower than graphene.~\citep{PhysRevB.76.064120,10.1063/1.4885215} 

Before we discuss the kinetics of point-defects, we investigate the formation of lattice vacancy in the light of recent experimental observations. The removal of a single P-atom introduces dangling electrons on the undercoordinated atoms, which concurrently undergo differential structural rearrangements to form two distinct defect configurations, SV(5$|$9)  and  SV(55$|$66) in Figure~\ref{fig:Fig1config}(a) and \ref{fig:Fig1config}(b). The SV(5$|$9) configuration involves inward displacement of two atoms along the armchair direction to form a new P--P bond between the two half-layers, while the third P-atom remains undercoordinated [Figure~\ref{fig:Fig1config}(a)]. In comparison, the vacancy relaxation in graphene results in multiple Jahn-Teller distorted configurations, where the planar (5$|$9) vacancy is the ground state.~\citep{PhysRevB.98.075439} The second configuration SV(55$|$66) in phosphorene undergoes minor displacement of atoms to form pentagon and hexagon pairs [Figure~\ref{fig:Fig1config}(b)].  The P--P bond along the armchair direction (2.57 \AA) is weaker than both pristine (2.27 \AA) and SV(5$|$9) defect with 2.41 \AA~ bonds.  Such four-fold coordination also appears in rippled graphene due to increased sp$^{3}$ character.~\citep{10.1021/jp300861m} The SV(5$|$9) vacancy is thermodynamically more stable than the SV(55$|$66) configuration [Figure~\ref{fig:Fig1config}(c) and Table~\ref{tablepvS1}]. Further, owing to a much lower $E^f$, the vacancy formation is much easier in phosphorene than in graphene ($\sim$ 7.5 eV)~\citep{PhysRevB.98.075439} and in mono-elemental bulk semiconductors such as Si and Ge ($E^f$ = 3.0--4.5 eV).~\citep{PhysRevB.88.195204,PhysRevB.87.035203} This indicates phosphorene to be defect-prone, and thus the point-defect driven structural degradation becomes relevant. Further, the vacancy formation energy at the surface of a few-layer phosphorene decreases with the increasing layer thickness  [Figure~\ref{fig:Fig1config}(c)].  While the relative stability of SV(5$|$9) and SV(55$|$66) differs by 350 meV in single-layer, it decreases to 48 meV in bulk BP. Similarly, the calculated $E^f$ for the subsurface SV(5$|$9)  and  SV(55$|$66) defects in 3L phosphorene is much lower, 1.54 and 1.60 eV, respectively, compared to the SLP (Table~\ref{tablepvS1}).  
  
\subsection{Anisotropic vacancy migration} 
A complete understanding of migration mechanisms of a lattice vacancy in phosphorene becomes important due to their significant presence owing to their low formation energies, and as they drive mechanical degradation.~\citep{10.1063/1.5016988} Here, we investigate both intra- and inter-layer diffusion and consider jumps along both the armchair and zigzag directions to understand the effects of structural anisotropy. Relating the degradation mechanisms with the vacancy migration along with the adatom mediated self-healing will assist in constructing a better defect annealing strategy.

\begin{figure*}[t]
\begin{center}
\includegraphics[width=0.75\textwidth]{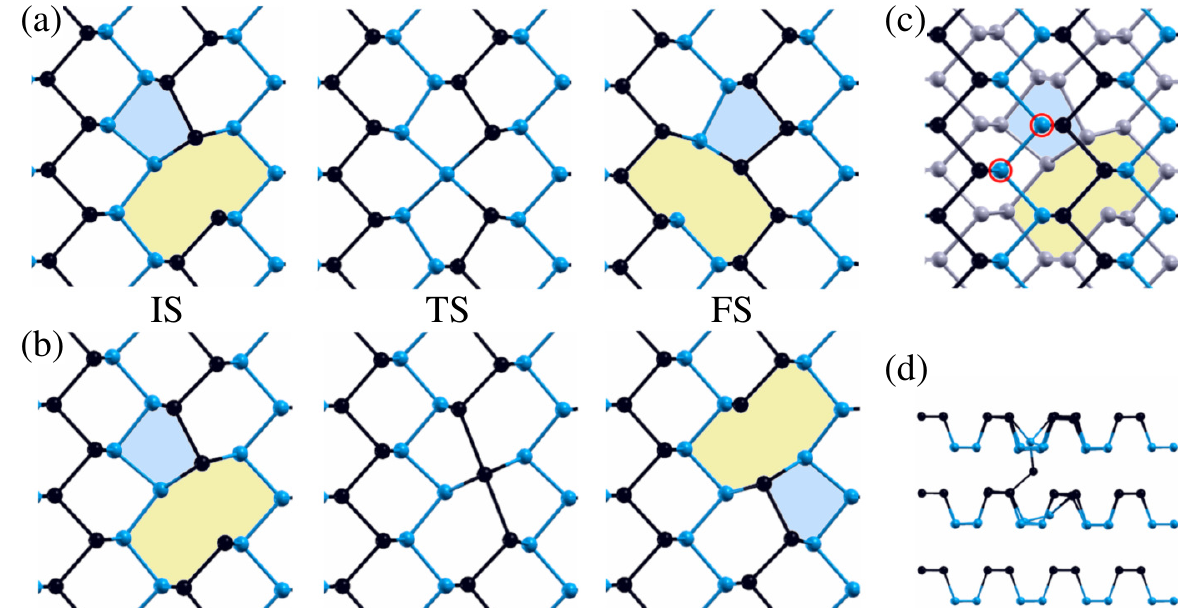}
\caption{The intra-layer SV(5$|$9) migration mechanisms along the (a) armchair and (b) zigzag crystallographic directions. The respective initial state (IS), transition state (TS) and final state (FS) are shown. The TS configurations are markedly different from each other, which leads to distinct activation barriers along the armchair and zigzag directions (Table~\ref{tablepvS1}). (c) The inter-layer SV(5$|$9) migration is studied by considering two different pathways, where either of the nearest and next-nearest neighbour atoms from the undefected half-layer jumps to the defected layer. The migrating atoms are encircled in red colour, and the grey atoms belong to the defected layer. (d) The side view showing the inter-layer P$-$P bonding (Wigner-like defect) for the corresponding transition state.}
\label{fig:Fig2NEB}
\end{center}
\end{figure*}

The SV(5$|$9) migration in the SLP is anisotropic along the armchair and zigzag directions with 0.38 and 0.12 eV, respectively [Table~\ref{tablepvS1} and Figure~\ref{fig:Fig2NEB}(a)-(b)]. Along the armchair direction the vacancy jump occurs via switching of bonds -- the P-atom of the pentagon bond is displaced along the armchair direction and subsequently bonds with the undercoordinated atom [Figure~\ref{fig:Fig2NEB}(a)]. The undercoordinated atom then migrates to the bottom half-layer in the final configuration. The vacancy migration along the zigzag direction occurs analogously shown in Figure~\ref{fig:Fig2NEB}(b); however, the migration is facile due to a much lower activation barrier. These results are in good agreement with the previously reported values of 0.18$-$0.40 eV.~\citep{C6NR05414E,0957-4484-26-6-065705,10.1021/jacs.6b04926}

In addition to the anisotropic migration, the vacancy jump mechanism discussed so far is peculiar since the undercoordinated atom plays a minor role in the migration. Typically, the undercoordinated atom can detach easily and jump to the vacant lattice site. However, during SV(5$|$9) vacancy migration, phosphorene retains the staggered arrangement by switching of P--P bonds across the half-layers and lowers the activation barrier. The overall diffusion occurs through a combination of anisotropic vacancy jumps, and the diffusion along the armchair direction is the rate-limiting step with 0.38 eV barrier.  In contrast, a relatively high barrier of 0.72 eV was calculated for the vacancy migration in graphene.~\citep{PhysRevB.98.075439} The aggregation of vacancy defects in an anisotropic environment will lead to the formation of line defects and grain boundaries, which in turn will act as a sink for diffusing vacancies.~\citep{C6NR05414E} 

\begin{figure}[!t]
\begin{center}
{\includegraphics[width=0.46\textwidth]{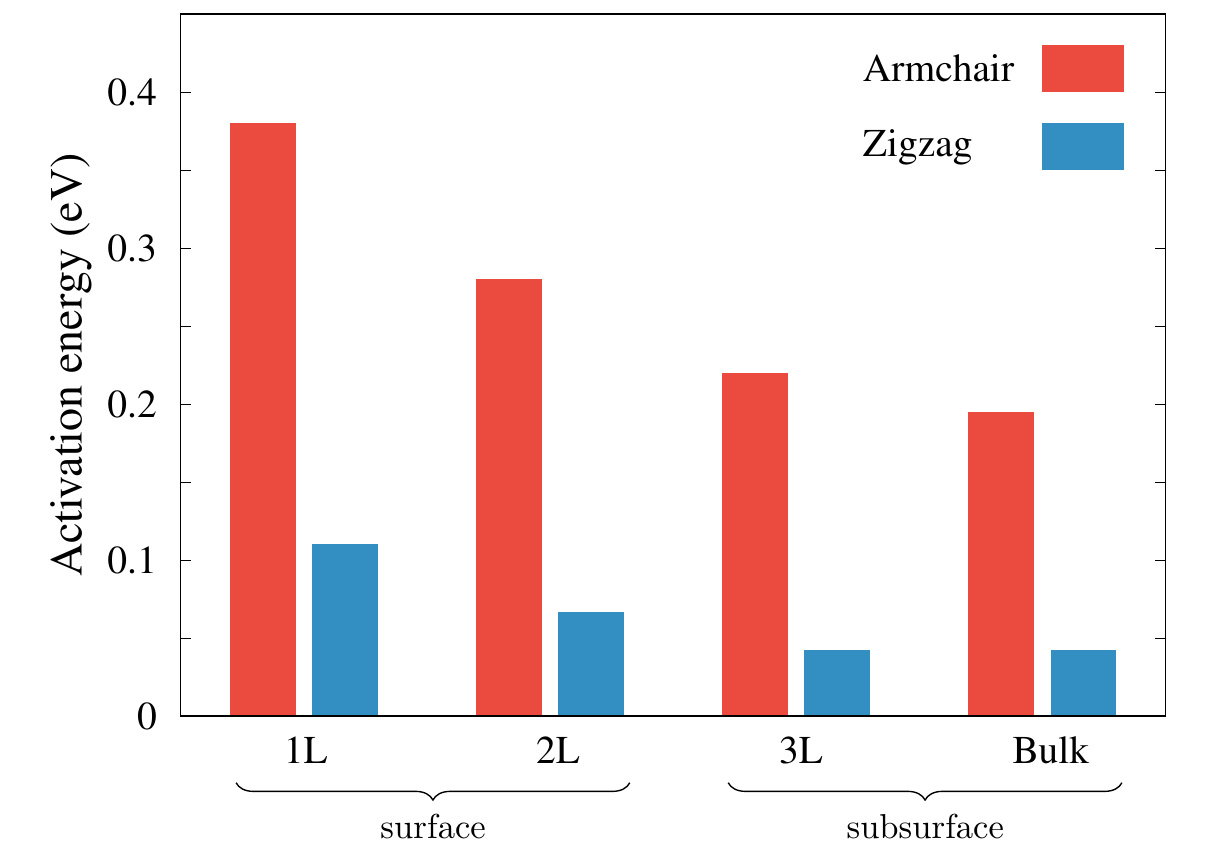}}
\caption{Activation barriers for the in-plane diffusion of SV(5$|$9) vacancy at the surface and subsurface. The anisotropic crystal structure of phosphorene results in two distinct migration pathways along the armchair and zigzag directions with different activation barriers. Calculated barriers for the surface vacancy depend on the layer thickness. Further, the barriers are much smaller for the subsurface migration compared to the vacancy at the surface.}
\label{fig:Fig3histo}
\end{center}
\end{figure}

Now we investigate the effect of layer thickness on the in-plane vacancy diffusion, and also discuss how the migration is affected while the vacancy is embedded in few-layer phosphorene (Figure~\ref{fig:Fig3histo}). In 2L phosphorene, the activation barrier decreases to 0.28 and 0.07 eV, respectively, for diffusion along the armchair and zigzag directions (Figure~\ref{fig:Fig3histo}). We do not anticipate further  reduction in $E_a$ with increasing thickness. While the mechanism remains the same, the subsurface vacancy diffusion is easier as revealed by the lower activation barrier (Figure~\ref{fig:Fig3histo}). For subsurface SV(5$|$9) migration in 3L phosphorene, the calculated $E_a$ are found to be 0.22 and 0.04 eV along the armchair and zigzag directions, respectively. Therefore, in 3L phosphorene, the $E_a$ for the subsurface diffusion converges to bulk BP, which we find to be 0.2 and 0.04 eV, respectively (Figure~\ref{fig:Fig3histo}). These results are in contrast to the few-layer graphene and graphite, where the activation barrier is higher for the subsurface vacancy diffusion,~\citep{PhysRevB.98.075439,PhysRevB.47.11143,10.1021/jp901578c} and the microscopic origin has been discovered only recently.~\citep{PhysRevB.98.075439} In single-layer graphene, the strain generated in the lattice due to the vacancy diffusion is released by an out-of-plane buckling, whereas the corresponding strain in phosphorene is released by the lattice relaxation within the half-layers. As the out-of-plane buckling is restricted in multi-layered graphene, the $E_a$ for the surface and subsurface vacancy increases substantially. In contrast, the lower $E_a$ in few-layer phosphorene is due to the reduction in the corresponding formation energies as discussed earlier [Figure~\ref{fig:Fig1config}(c)] ~\citep{PhysRevB.91.045433,10.1038/srep14165} A similar Br{\o}nsted-Evans-Polanyi-type correlation between $E^f$ and $E_a$ has been reported in other bulk semiconductors and carbon nanotubes.~\citep{10.1063/1.2335842,PhysRevB.79.092101, acs.jpcc.5b11682}

While the $E^f$ for the subsurface SV(5$|$9) defect is lower than the same on the surface, it would be interesting to investigate the subsurface vacancy diffusion to the surface. Due to the AB-staking in few-layer phosphorene, the vacancies in adjacent layers do not align vertically and thus diffuses via inter-layer diagonal jumps [Figure~\ref{fig:Fig2NEB}(c)]. We considered the two possible pathways, where either of the nearest and next nearest neighbour atom from the adjacent undefective layer migrates to the SV(5$|$9) vacancy. The migrating P-atom, in both mechanisms, forms a Wigner-like defect, where the moving P-atom bonds with both active layers [Fig.~\ref{fig:Fig2NEB}(d)].~\citep{Telling2003} While the inter-layer SV(5$|$9) migration via the nearest P-atom jump is calculated to be 2.17 eV, the second mechanism costs much higher activation energy of 2.80 eV. Such substantial inter-layer barrier restricts the vacancy migration across the layers to very high-temperature and does not play any significant role in the annealing process, which is usually performed at temperatures around 475 K.~\citep{10.1021/acs.jpclett.5b00043} In comparison, the corresponding barrier is significantly higher in few-layer graphene, 5.5--7.0 eV, and becomes relevant only above 2000 K.~\citep{PhysRevB.90.174108,C4NR00488D}  However, the inter-layer vacancy migration may occur at a much lower temperature in the presence of complex inter-layer defect structures. The similar increase in inter-layer activity is reported in the few-layer phosphorene above 600 K, while such inter-layer defect complexes are present.~\citep{0957-4484-29-6-065702}

\subsection{Vacancy-driven degradation, void formation and sublimation} 
The recent experiments indicate that heating of few-layer phosphorene creates oblate voids in the individual layers, which is followed by the inter-layer P--P bond formation. Further, the degradation proceeds via amorphization and eventual sublimation between 650--680 K.~\citep{10.1021/acs.jpclett.5b00043,0957-4484-29-6-065702,10.1021/acs.jpclett.6b00584} Thus, to achieve a better understanding of structural degradation, we investigate a cohort of mechanisms such as self-healing, [SV + SV] $\rightarrow$ DV, SV $\rightarrow$ [DV + P$_{\rm I}$] pair transformations, and sublimation (Table~\ref{tablepvS1}).

\begin{figure}[!t]
\begin{center}
\includegraphics[width=0.48\textwidth]{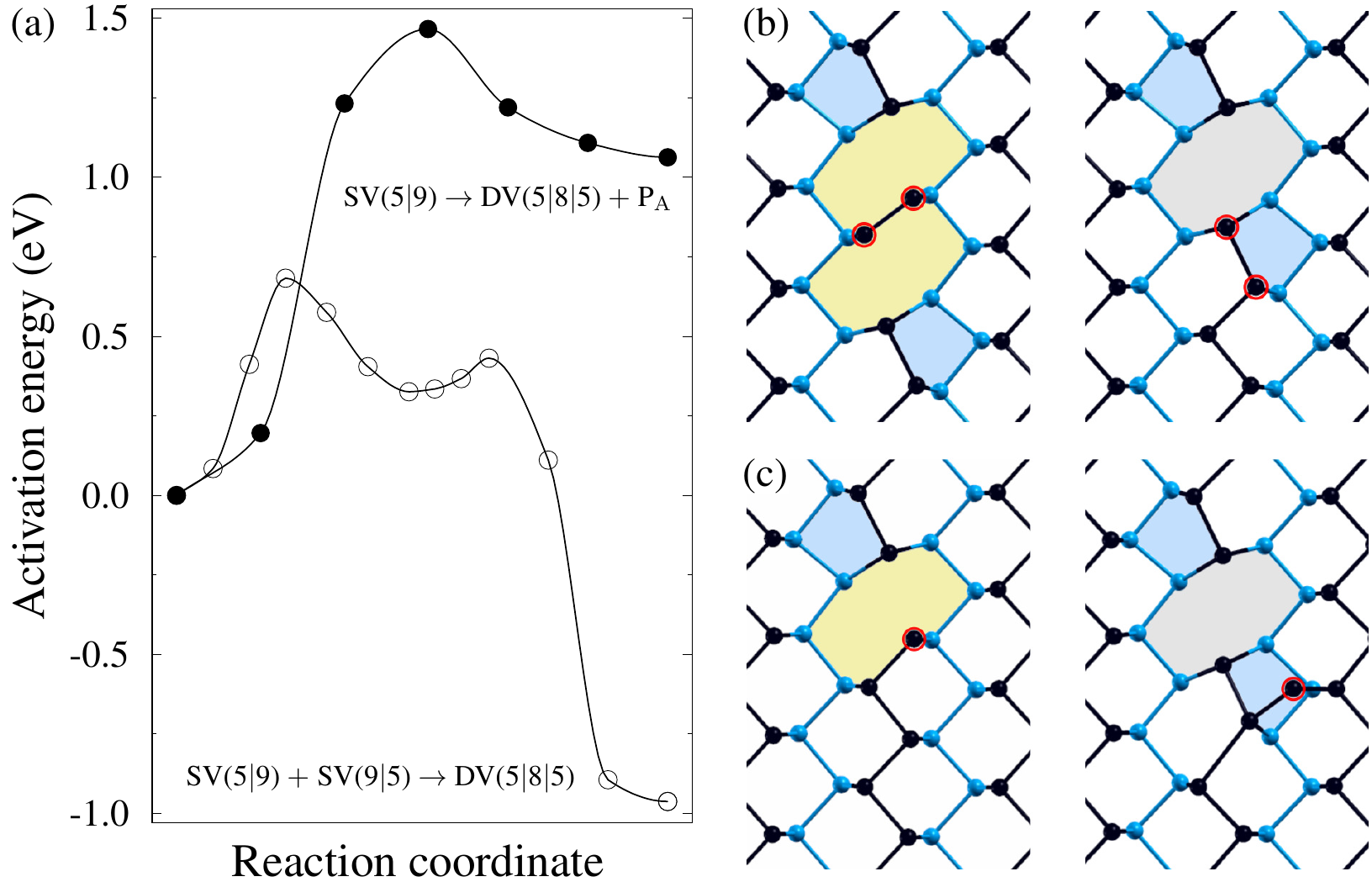}
\caption{
(a) The energy profiles for the mechanisms leading to the larger defect formation, SV(5$|$9) + SV(9$|$5) $\rightarrow$ DV(5$|$8$|$5)  and  SV(5$|$9) $\rightarrow$ DV(5$|$8$|$5) + P$_{\rm A}$. Owing to a lower 0.68 eV barrier, DV(5$|$8$|$5) is generated via coalescence of itinerant SV(5$|$9) vacancies. While the forward SV(5$|$9) $\rightarrow$ DV(5$|$8$|$5) + P$_{\rm A}$ process has a high barrier, note that the reverse self-healing process has a much lower barrier of 0.46 eV. (b)-(c) The corresponding initial and final structures. (b) The merger of mono-vacancies into DV(5$|$8$|$5) through bond rotation (encircled in red). (c) The structures for the SV(5$|$9) $\rightarrow$ DV(5$|$8$|$5) + P$_{\rm A}$ mechanism. 
}
\label{fig:Fig3bMV2DV}
\end{center}
\end{figure} 

The divacancy defect DV(5$|$8$|$5) is formed by the removal of two adjacent P-atoms from the same half-layer. The concurrent structural relaxation to pentagon-octagon-pentagon arrangement ensures three-fold coordination for all the atoms with a formation energy of 1.56 eV, which is lower than the SV(5$|$9) defect formation. Although the DV(5$|$8$|$5) has not been yet observed in the low-temperature experiments below 10 K,~\citep{10.1021/acs.nanolett.7b00766,10.1063/1.5016988} it should be formed during exfoliation or by the coalescence of two itinerant SV(5$|$9) defects. Further, the [SV(5$|$9) + SV(9$|$5)] $\rightarrow$ DV(5$|$8$|$5) merger (Supplemental Material~\citep{supple}) is viewed as one of the relevant mechanisms for multi-vacancy defect formation, and the calculated barrier is found to be 0.68 eV [Table~\ref{tablepvS1} and Figure~\ref{fig:Fig3bMV2DV}(a)]. The corresponding mechanism is shown in the Figure~\ref{fig:Fig3bMV2DV}(b). The reverse mechanism of DV(5$|$8$|$5) splitting into two adjacent vacancies requires much higher energy of 1.64 eV. Thus, the itinerant SV defects will easily coalesce to form DV(5$|$8$|$5) and larger defects. In contrast to the present calculations, a much smaller barrier of 1.05 was estimated earlier for the DV(5$|$8$|$5) $\rightarrow$ [SV(5$|$9) + SV(9$|$5)] dissociation.~\citep{10.1021/jacs.6b04926} The discrepancy arises since the previous estimate is based on the energy differences between the DV and [SV + SV] structures only and without the kinetic consideration of bond reorientation. The split vacancy may also diffuse away, along the armchair and zigzag directions, with activation barriers that decrease with increasing the distance between them. The calculated $E_a$ converges to the isolated vacancy migration, while the vacancies are sufficiently separated.~\cite{supple} While compared with the single-layer graphene, the DV defects are known to be stable due to a much high barrier for dissociation above 5 eV.~\citep{PhysRevLett.95.205501,10.1021/jp405130c} 

At present, the proposed mechanism for the anisotropic void formation, that is experimentally observed above 650 K, is the removal of edge atoms from the defects.~\citep{10.1021/acs.jpclett.5b00043,0957-4484-29-6-065702,10.1021/acs.jpclett.6b00584} However, the corresponding rate-determining atomistic process is not yet known. In this regard, we investigate the transformation of an SV(5$|$9) into the [DV(5$|$8$|$5) + P$_{\rm A}$] complex, where the under-coordinated P-atom of the SV(5$|$9) migrates to the nearest bridge site [Figure~\ref{fig:Fig3bMV2DV}(c)]. An activation barrier of 1.46 eV is calculated for this mechanism [Table~\ref{tablepvS1} and Figure~\ref{fig:Fig3bMV2DV}(c)], which is in excellent agreement with the experimental estimate of 1.64 $\pm$ 0.1 eV.~\citep{10.1021/acs.jpclett.6b00584} Subsequent removal of the peripheral atoms will require a similar or lower barrier due to weak bonding at the edges and will result in void formation. Further, the P$_{\rm A}$-adatoms generated in the above process will diffuse along the zigzag direction with extremely low migration barrier of 0.19 eV, which is much higher 1.05 eV along the armchair direction (Table~\ref{tablepvS1}).

As the temperature is further increased above 623 K, the degradation process is accelerated, and results in rapid sublimation between 650--680 K.~\citep{10.1021/acs.jpclett.5b00043,0957-4484-29-6-065702,10.1021/acs.jpclett.6b00584} We address the sublimation process by considering the desorption of single and paired P-adatoms that are produced during the SV(5$|$9) $\rightarrow$ [DV(5$|$8$|$5) + P$_{\rm A}$] process. The single P$_{\rm A}$ on the pristine surface binds at the bridge site connecting the two armchair edges with 1.43 eV formation energy. The shorter adatom-surface separation of 2.19 \AA~ highlights the role of lone-pair electrons in the bonding. The desorption mechanism is modelled by vertically detaching the P$_{\rm A}$ by 6 \AA~ from the surface, which requires overcoming a minimum barrier of 2.14 eV (Table~\ref{tablepvS1}). 

Highly itinerant P$_{\rm A}$ on the surface forms P$_{\rm A}$-pairs, and we investigate the simultaneous desorption of both P-adatoms. Two adjacent P$_{\rm A}$ (Supplemental Material~\citep{supple}) along the zigzag directions is found to be thermodynamically most favourable with 2.01 eV formation energy. Thus, the pair formation is thermodynamically advantageous than the two non-interacting P$_{\rm A}$s, which favours aggregation of adatoms. Further, the desorption of P$_{\rm A}$-pair is found to be a spontaneous process without an activation barrier, and the free-energy is reduced by 0.82 eV/P$_{\rm A}$-pair.  Therefore, we conclude that the itinerant P$_{\rm A}$s thermodynamically form P$_{\rm A}$-pairs, and its concurrent desorption is critical to the rapid sublimation of phosphorene. These results are in excellent agreement with the recent experimental hypothesis of pairwise sublimation.~\citep{10.1021/acs.jpclett.6b00584}   

We argue that the origin of anisotropic void formation is two-fold -- directional anisotropy in both vacancy and adatom diffusion. Faster vacancy diffusion along the zigzag direction triggers multi-vacancy defects that are elongated along the zigzag direction. Further, the faster adatom diffusion along the zigzag path produces P$_{\rm A}$-pairs promoting sublimation. Thus, these processes will create anisotropic voids with the long-axis along the zigzag direction, which is in agreement with the recent experimental observations.~\citep{10.1021/acs.jpclett.6b00584, aadd20}

\begin{figure}[!t]
\begin{center}
{\includegraphics[width=0.48\textwidth]{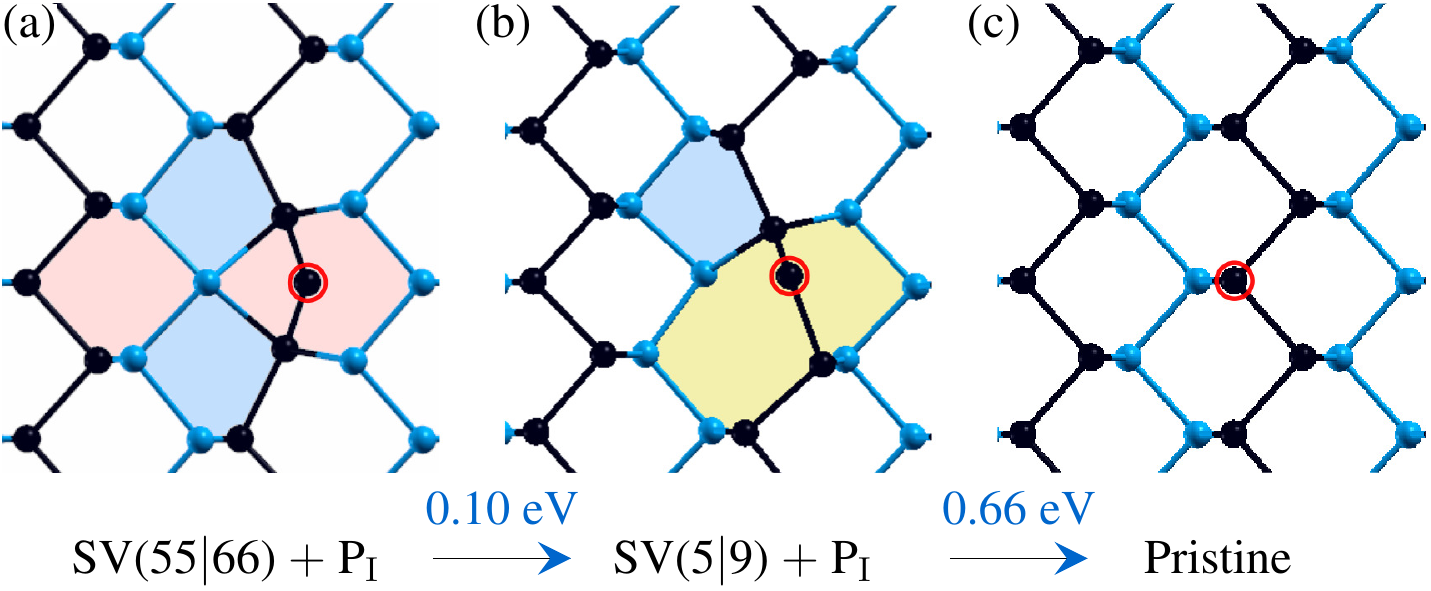}}
\caption{Self-healing mechanisms of a vacancy via the recombination with P interstitial P$_{\rm I}$ (encircled by red colour), which are rate-limited by 0.69 eV activation barrier [Table~\ref{tablepvS1}]. Frenkel defects (a) SV(55$|$66) + P$_{\rm I}$ and (b) SV(5$|$9) + P$_{\rm I}$ recombine to form pristine lattice in (c). The numbers in blue colour indicate the corresponding activation barrier. The interstitial P$_{\rm I}$ in the Frenkel defects preferentially binds along the zigzag direction, and these complexes are thermodynamically stable at room-temperature. The DV(5$|$8$|$5) is partially healed first to an SV(5$|$9) defect by absorbing an itinerant P$_{\rm A}$ adatom, which requires 0.46 eV energy (Table~\ref{tablepvS1} and Figure~\ref{fig:Fig3bMV2DV}).}
\label{fig:Fig4Ad}
\end{center}
\end{figure}

\subsection{Defect healing and the role of P-adatom} 
Point-defects can generate larger defects via various mechanisms that are discussed above. However, the point-defects can also be self-healed by interacting with the adatom or interstitial defects, which should be investigated in detail. The vacancy-interstitial (P$_{\rm I}$) Frenkel pairs with SV(5$|$9) and SV(55$|$66) vacancies [Figure~\ref{fig:Fig4Ad}(a) and \ref{fig:Fig4Ad}(b)] are stable with   1.56 and 1.93 eV formation energies, respectively. In comparison, such Frenkel defects are absent in graphene due to the dominant nature of sp$^{2}$ bonding, which prevents out-of-plane geometry of the  C-adatom within the graphene vacancy. The self-healing of SV(5$|$9) by interstitial P$_{\rm I}$ [Figure~\ref{fig:Fig4Ad}(b) $\rightarrow$ \ref{fig:Fig4Ad}(c)] requires a moderate $E_a$ of 0.69 eV. In contrast, the relatively less stable SV(55$|$66) and interstitial pair [Figure~\ref{fig:Fig4Ad}(a)] first converts into an intermediate [SV(5$|$9) + P$_{\rm I}$] pair [Figure~\ref{fig:Fig4Ad}(b)] with 0.1 eV activation energy (Table~\ref{tablepvS1}). In contrast, the reverse mechanism of [SV(5$|$9) + P$_{\rm I}$] Frenkel pair generation at the pristine lattice requires substantially higher energy of 2.25 eV (Table~\ref{tablepvS1}).

Having studied the self-healing of single-vacancy, we now investigate the DV(5$|$8$|$5) healing. The partial healing of DV(5$|$8$|$5) to SV(5$|$9) by an itinerant P$_{\rm A}$ requires an activation barrier of 0.46 eV [Table~\ref{tablepvS1} and Figure~\ref{fig:Fig3bMV2DV}(c)], and this process is accessible at 300 K [Fig.~\ref{fig:Fig4Ad}(d--e)]. The complete healing of DV then proceeds via the absorption of another mobile P$_{\rm A}$ into the SV with 0.69 eV barrier as discussed earlier, which is the rate-limiting process and occurs at a higher temperature above 450 K.~\citep{10.1021/acs.jpclett.5b00043} In contrast to phosphorene, this process requires 0.9 eV of energy in graphene and thus, the adatom-DV(5$|$8$|$5) complex is experimentally observed at room-temperature.~\citep{10.1038/nature02817,PhysRevLett.100.175503,10.1021/jp405130c}

\begin{figure}[!t]
\begin{center}
\includegraphics[width=0.48\textwidth]{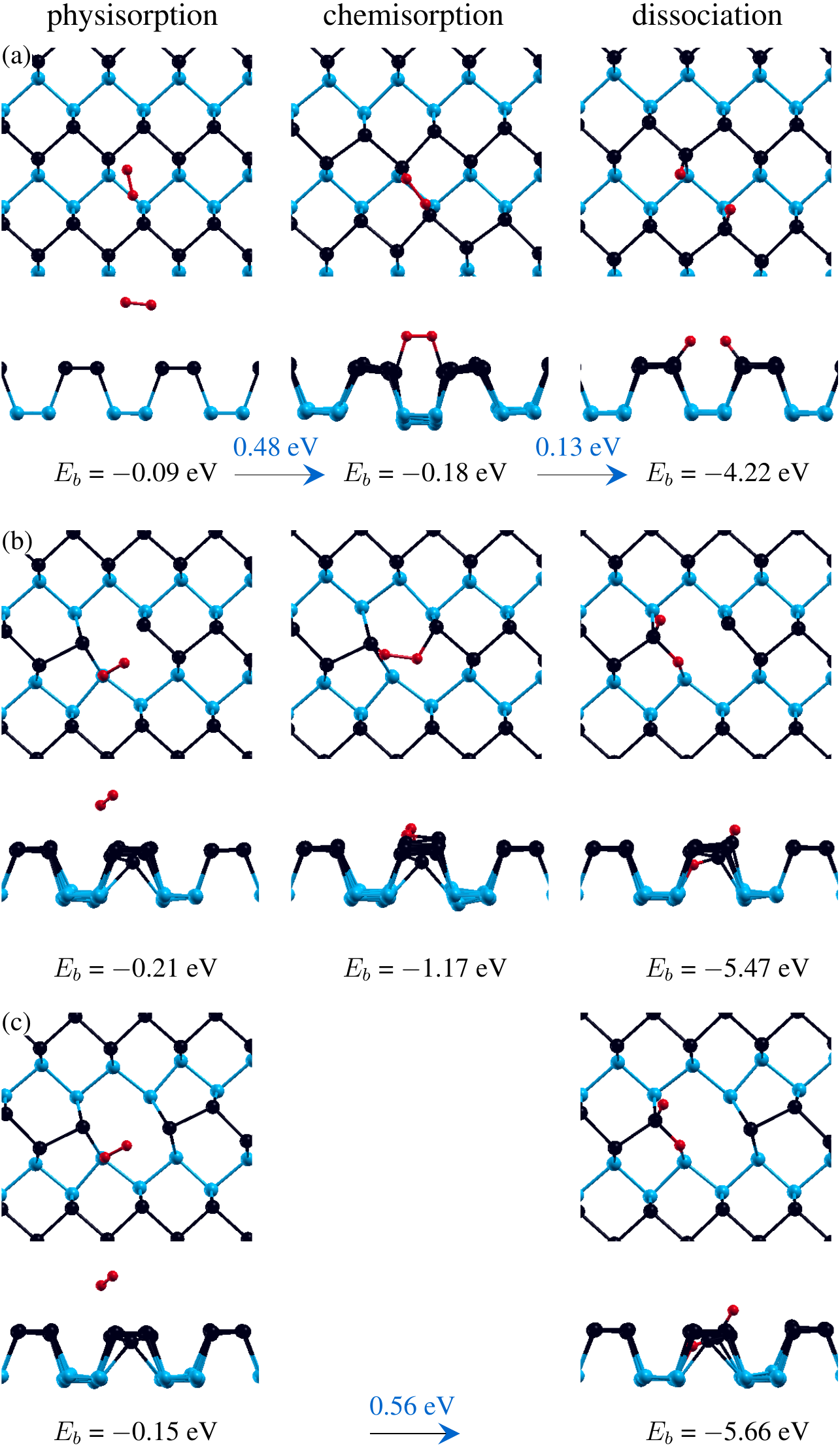}
\caption{The chemical degradation of phosphorene through oxidation. The top and side views for O$_{2}$-dissociation are shown (a) on the pristine surface; and at the (b) SV(5$|$9) and (c) DV(5$|$8$|$5) defect sites.   While the entire physisorption $\rightarrow$ chemisorption $\rightarrow$ dissociation process is activated on the pristine surface with 0.48 eV rate-limiting barrier, the same is spontaneous at the SV(5$|$9) vacancy.  We did not find any chemisorbed configuration at the DV(5$|$8$|$5) defect, and the O$_{2}$-dissociation directly proceeds from the physisorbed structure with 0.56 eV barrier. For all cases, the dissociated O-atoms bind strongly with the phosphorene, and would thus be impossible to remove from the lattice. The numbers in blue colour indicate the corresponding activation barrier.
}
\label{fig:Fig6Oxy}
\end{center}
\end{figure}

\subsection{Oxidation of pristine and defected phosphorene} 
In addition to the defect-mediated structural degradation, phosphorene also undergoes chemical degradation in the ambient environment. In this regard, we investigate the O$_2$-dissociation mechanism on the pristine and defected phosphorene (Supplemental Material).~\cite{supple}  The binding energy is calculated as $E_b$ = $E_{\rm (SLP-O_2)} - E_{\rm SLP} - E_{\rm O_2}$, where $E_{\rm SLP}$ is the energy of the pristine SLP, $E_{\rm O_2}$ is the energy of the O$_2$-molecule and $E_{\rm (SLP-O_2)}$ is the energy of the interacting composite system. The physisorbed O$_{2}$-molecule is very weakly bound ($E_b$ = $-$0.09 eV) to the SLP at 3.17 \AA~ vertical height [Figure~\ref{fig:Fig6Oxy}(a)].~\citep{supple} The spin-state of the O$_2$-molecule plays a crucial role in the activated chemisorption and concurrent dissociation. The O$_{2}$-molecule undergoes a triplet to singlet spin conversion ($S=1$ $\rightarrow$ $S=0$) with increasing proximity to phosphorene. The chemisorption in the singlet state is an activated process with 0.48 eV barrier [Figure~\ref{fig:Fig6Oxy}(a)], which is in good agreement with the previous calculation.~\citep{PhysRevLett.114.046801}  In this chemisorbed state, the O$_{2}$ is still weakly bound to the SLP ($E_b$=$-$0.18 eV), while the O-atoms are weakly bonded to the P-atoms across the armchair direction with an activated O$-$O bond. The ultimate dissociation of the chemisorbed O$_{2}$ requires 0.13 eV barrier, and the dissociated O-atoms form strong dangling P$-$O bonds with $-$4.22 eV binding energy [Figure~\ref{fig:Fig6Oxy}(a)].

It would be interesting now to investigate the O$_2$-dissociation in the presence of lattice defects.  The physisorption at a 2.06 \AA~ height above the SV(5$|$9) vacancy is still weak with $-$0.21 eV binding energy [Fig.~\ref{fig:Fig6Oxy}(b)]. Similar to the pristine SLP, the spin-state of the O$_2$ is $S=1$. Remarkably, we find that the entire physisorption $\rightarrow$ chemisorption $\rightarrow$ dissociation process is spontaneous without any activation barrier. The present results are in contrast to the previous study,~\citep{2053-1583-4-1-015010} where a 0.59 eV barrier was predicted due to a different physisorbed configuration, which we find to be a metastable state in the present calculations. The chemisorbed structure has a significantly higher binding energy of $-$1.17 eV compared to the same on the pristine phosphorene due to a strong P$-$O bond that is accompanied by an increased puckering at the SV site [Figure~\ref{fig:Fig6Oxy}(b)]. Unlike the case of pristine SLP, the dissociated O-atoms embed within the lattice and results in a large gain in binding energy.~\citep{supple} A similar triplet to singlet spin-conversion takes place during the chemisorption at the SV(5$|$9) vacancy. While the O$_2$-dissociation process is similar at the SV(55$|$66) vacancy, the only difference is that the physisorption $\rightarrow$ chemisorption process requires 0.53 eV activation energy. In comparison, we did not find the existence of a chemisorbed state at the DV(5$|$8$|$5), and the direct physisorption $\rightarrow$ dissociation process proceeds with a 0.56 eV barrier [Figure~\ref{fig:Fig6Oxy}(c)]. In all cases, it is understood that once O$_2$ is completely dissociated, it would be difficult to remove from the lattice due to strong P$-$O binding.   


\begin{figure*}[!t]
\begin{center}
{\includegraphics[width=0.90\textwidth]{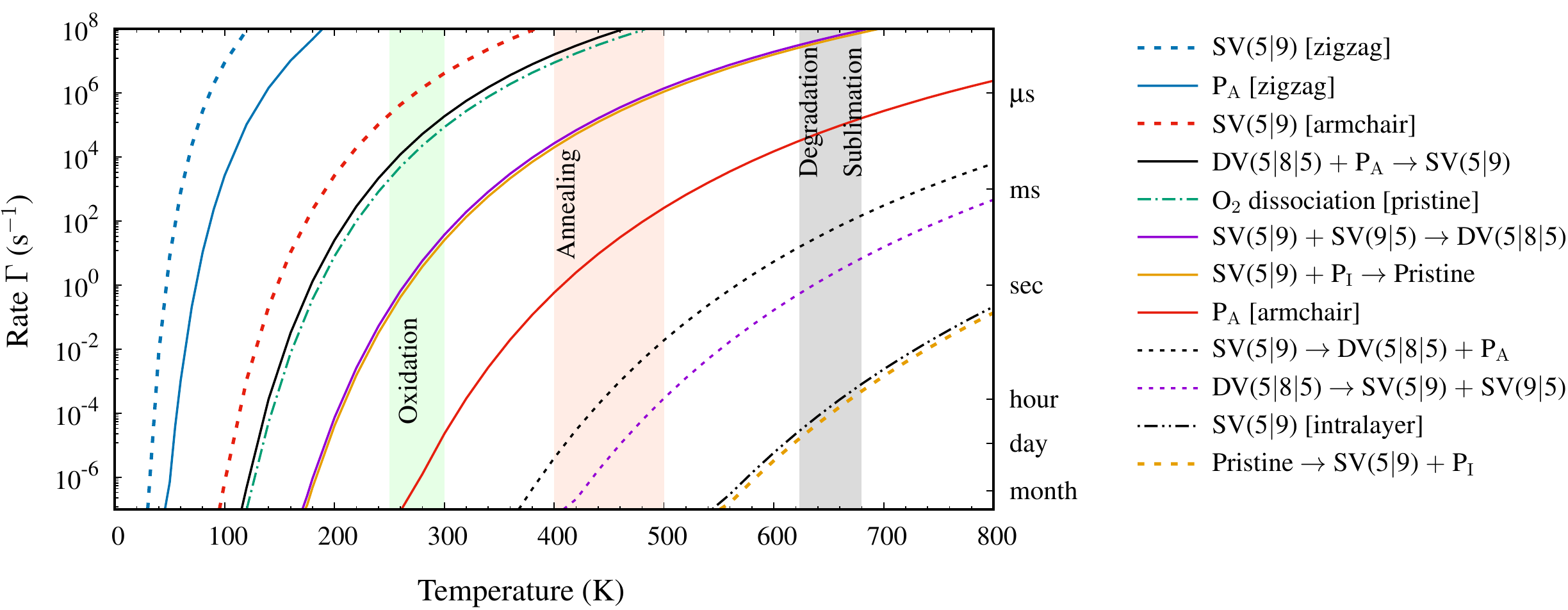}}
\caption{Arrhenius rate $\Gamma$ calculated for the various microscopic mechanisms leading to oxidation, annealing, degradation and sublimation.  A typical pre-factor of 10$^{13}$ s$^{-1}$ is assumed. The SV(5$|$9) and P$_{\rm A}$ are highly itinerant even below 100 K, which triggers the competing self-healing of vacancy and itinerant SV(5$|$9) mergers above 450 K. The degradation is accelerated above 650 K through the emission and sublimation of P$_{\rm A}$. The chemical degradation through oxidation of  pristine SLP takes place above 250 K. The overall phase-diagram are in excellent agreement with the recent experimental observations.~\citep{10.1021/acs.jpclett.5b00043,0957-4484-29-6-065702,10.1021/acs.jpclett.6b00584}}
\label{fig:Fig5jump}
\end{center}
\end{figure*}

\subsection{Degradation phase diagram}
The defect formation energies and the kinetic pathways discussed so far (Table~\ref{tablepvS1}) provide a comprehensive insight into the defect thermodynamics  and their concurrent microscopic mechanisms leading to self-healing, oxidation, degradation and sublimation at relevant temperatures. To understand the temperature dependence of the various microscopic mechanisms, we calculate the kinetic rate and temperature phase diagram (Figure~\ref{fig:Fig5jump}) using the Arrhenius equation $\Gamma = \nu_0 \exp(-E_a/k_BT)$. The pre-factor $\nu_0$ is the Debye frequency, which is typically 10$^{13}$ s$^{-1}$.

At temperatures below 100 K, both the SV(5$|$9) and P$_{\rm A}$ defects are highly itinerant along the zigzag direction, while all other kinetic processes are hindered (Figure~\ref{fig:Fig5jump}). However, the anisotropic SV(5$|$9) diffusion in the armchair direction within measurable time-scale occurs only above 150 K. In contrast, the P$_{\rm A}$ diffusion along the armchair direction of the phosphorene lattice is predicted to take place at a much higher temperature above 450 K.  

Self-healing of the vacancy defects are rate-limited by the merger of SV(5$|$9) with P$_{\rm I}$, which appears above 400 K (Figure~\ref{fig:Fig5jump}), which is in agreement with the experimental observation.~\citep{10.1021/acs.jpclett.5b00043} It is important to note that at this temperature range, the itinerant SV(5$|$9) vacancies also merge to form bigger vacancy defects. Thus, we predict a complex interplay between the competing mechanisms leading to self-healing and extended defect formation. However, phosphorization above 400 K will anneal the vacancy defects. 

Further, structural degradation will be accelerated beyond 650 K. In addition to much faster SV(5$|$9) mergers,  the larger vacancy defects are also created from the SV(5$|$9) point-defects through simultaneous emissions of P$_{\rm A}$. The generated P$_{\rm A}$s concurrently form P$_{\rm A}$-pairs via diffusion and are detached spontaneously as P$_{\rm A}$-pair from the surface. Thus, above 650 K, the rapid pair-wise sublimation occurs, and the overall degradation mechanism is in agreement with the recent experimental observations.~\citep{10.1021/acs.jpclett.5b00043,0957-4484-29-6-065702,10.1021/acs.jpclett.6b00584} We also find that below 1000 K temperature, the inter-layer vacancy diffusion does not play any role in degradation in few-layer phosphorene. The chemical degradation via oxidation takes place above 250 K (Figure~\ref{fig:Fig5jump}), and after that, it is impossible to remove the dissociated O-atoms from the lattice.

\section{Summary}
We investigate the structural and chemical degradation of phosphorene within the first-principles calculations. A cohort of microscopic mechanisms is studied to develop a degradation phase diagram. The vacancy diffusion is easily accessible below the room-temperature, which leads to their merger into larger vacancy defects above 400 K. The bond-rotation mechanism of two neighbouring SV(5$|$9) to form DV(5$|$8$|$5) is found to be the rate-limiting mechanism. The self-healing of vacancy via itinerant adatom absorption is also triggered at a similar temperature range. In addition to the merger of mobile vacancies, the emission of the undercoordinated P-atom from the point-defects also generates two-dimensional anisotropic voids above 650 K. Such P-atoms are highly itinerant and thermodynamically form P$_{\rm A}$$-$P$_{\rm A}$-dimer, and further degradation proceeds through the spontaneous pair-wise P$_{\rm A}$ sublimation. While the inter-layer vacancy diffusion in few-layer phosphorene is mostly blocked owing to the high activation barrier, the merger of sub-surface vacancies is much faster at a given temperature compared to the same at the surface. However, the degradation through P-emission from the point-defect and the competing self-healing through the P-diffusion are understandably blocked at the sub-surface level. Therefore, the degradation activity including oxidation primarily occurs at the surface. 

The chemical degradation that is observed in ambient condition proceeds via O$_2$-dissociation, which we find to be accessible at the room-temperature on the pristine surface, and notably the process is spontaneous at the single-vacancy site. Further, it is impossible to remove the dissociated O-atoms from the lattice, and the most desirable electronic properties of phosphorene are permanently lost. Thus, the present work provides with the microscopic insights into the phosphorene degradation, that will have similarities with the other puckered layered two-dimensional materials.

%
%

\section{Acknowledgements}
M.K. acknowledges funding from the Department of Science and Technology through Nano Mission project SR/NM/TP-13/2016, and the Science and Engineering Research Board through EMR/2016/006458 grant. The computational facilities at the Centre for Development of Advanced Computing, Pune; and at the Inter University Accelerator Center, Delhi; and at the Center for Computational Materials Science, Institute of Materials Research, Tohoku University are greatly acknowledged.


%

\end{document}